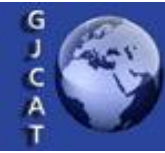

# Improved Robust DWT-Watermarking in YCbCr Color Space

**Atefeh Elahian[#1], Mehdi Khalili[#2], Shahriar Baradaran Shokouhi[#3]**

[#1]Electronic Learning Center,, University of Science & Technology,Teharan, Iran
[#2] Institute for Informatics and Automation Problems, National Academy of Science,Yerevan, Armenia
[#3] School of Electrical Engineering, University of Science & Technology,Teharan, Iran

*Abstract*—Digital watermarking is an effective way to protect copyright. In this paper, a robust watermarking algorithm based on wavelet transformation is proposed which can confirm the copyright without original image. The wavelet transformation technique is effective in image analyzing and processing. Thus the color-image watermark algorithm based on discrete wavelet transformation (DWT) begins to draw an increasing attention. In the proposed approach, the watermark Encrypt by Arnold transform and the host image is converted into the YCbCr color space. Then its Y channel decomposed into wavelet coefficients and the selected approximation coefficients are quantized and then their least significant bit of the quantized coefficients is replaced by the Encrypted watermark using LSB insertion technique. The experimental results show that watermark embedded by this algorithm is of better robustness and extra imperceptibility and robustness against wavelet compression compared to the traditional embedding methods in RGB color space.

## I. INTRODUCTION

With the rapid development of information technology and computer networks, the access and transmission of digital multimedia information become very convenient. But it also brings a lot of negative effects, such as information stealing, piracy, etc. [1]. Digital Watermarking is a method of embedding some identification information directly into digital data carrier such that it does not affect its original value and to the people's perceptual system it not easy to be aware of through the information hidden in the data carrier. We can confirm that the content creators, buyers and send secret information or to determine whether the data carrier has been tampered with or other purposes. Digital watermarking is an important research direction in technology of information hiding [2]. In general, we can classify digital watermark into two classes depending on the domain of watermark embedding, i.e., the spatial domain and the transform domain, where the properties of the underlying domain can be exploited. Previous works have shown that the transform domain scheme is typically more robust to noise, common image processing, and compression when compared with the spatial transform scheme [3]. According to the need for original data during the watermark detection process, digital watermark can be also classified into private and public (or blind) algorithms [4]. In this paper, an algorithm of digital image watermarking based on discrete wavelet transform is proposed. The fundamental advantage of our wavelet-based technique lies in the method used to embed the watermark in Y channel DWT approximation coefficients of YCbCr color space of host image using LSB insertion technique and using Arnold transform map that can eliminate spatial correlation and disperse the error bits among all pixels to make watermarking more strongly robust against cropping operation [5].

## II. ARNOLD TRANSFORM MAP

Watermark must be encrypted before embedding into its carrier. It will go through scrambling transformation so that the spatial correlation of watermark image pixels will be cancelled and its security will be strengthened. In this way, attackers cannot accurately identify the specific content of watermark even if they already extract it. A binary image after digital processing can be viewed as a matrix, one pixel corresponding to one matrix element. After linear or nonlinear transformation of the pixels in the matrix, the image will look desultorily. A binary watermark image will look more like noise after being transformed several times, thus, attackers will take it as noise and ignore it even when they know the imbedding algorithm and already extract the embedded data. Therefore, the security of watermark is strengthened. There are many common ways to scramble watermark images as a pre-treatment such as Arnold transformation, magic transformation, Hilbert curve, Conway game, broad Gray code transformation, affine transformation and orthogonal Latin square transformation [6]. In this paper, we adopt a simple way but with strong security Arnold transformation. Arnold transformation is posed in the research of Arnold ergodic theory, which is also called cat face transformation. The Arnold transformation is defined as follows:

$$\begin{bmatrix} x' \\ y' \end{bmatrix} = \begin{bmatrix} \begin{bmatrix} 1 & 1 \\ 1 & 2 \end{bmatrix} \cdot \begin{bmatrix} x \\ y \end{bmatrix} \end{bmatrix} \bmod N \quad (1)$$

where (x,y) is the location coordinates of the original image pixels, and (x',y') is the location coordinates of image pixels that after transform.

### III. YCbCr Color Space

YCbCr color space is used for color images cryptography. In this color space, the Y denotes the luminance component. It means that Y shows the brightness (luma). Also both of Cb and Cr represent the chrominance actors. It means that Cb is blue color minus luma (B_Y) and Cr is red color minus luma (R_Y) [7]. The difference between YCbCr color space and RGB color space is that YCbCr color space represents color as brightness and two color difference signals, while RGB represents color as red, green and blue. Equations 2 and 3 show transformations between RGB color space and YCbCr color space [8]:

$$\begin{bmatrix} Y \\ Cb \\ Cr \end{bmatrix} = \begin{bmatrix} 0.29890 & 0.58660 & 0.11450 \\ -0.16874 & -0.33126 & 0.50000 \\ 0.50000 & -0.41869 & -0.8131 \end{bmatrix} \begin{bmatrix} R \\ G \\ B \end{bmatrix} \quad (2)$$

$$\begin{bmatrix} R \\ G \\ B \end{bmatrix} = \begin{bmatrix} 1 & 0 & 1.40200 \\ 1 & -0.34414 & -0.71414 \\ 1 & 1.7720 & 0 \end{bmatrix} \times \begin{bmatrix} Y \\ Cb \\ Cr \end{bmatrix} \quad (3)$$

### IV. Discrete Wavelets Transform

Discrete wavelets transform is a brand-new signal analysis theory which has arisen in recent years. It is a new time and frequency domain analysis method which can localize time and frequency domain, and has widely used in many fields. The basic idea of DWT is the detailed frequency separation on signal, namely multi-resolution decomposition. The host graph is decomposed to four sub-graphs in size of one quarter: one low frequency approximation graph and three medium and high frequency detail sub-graphs in horizontal, vertical and diagonal direction. The human visual system (HVS) is more sensitive to image modifications in smooth areas (low frequency components) than texture and edges areas (high frequency components) [9]. Figure 1 is a schematic diagram on three-level decomposition of discrete wavelets.

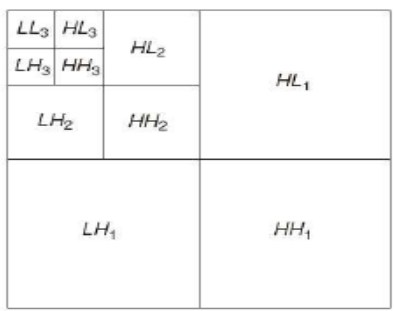

Fig. 1 Wavelet decomposition schematic diagram

### V. Watermarking Framework

The current study task of digital watermarking is to make watermarks invisible to human eyes as well as robust to various attacks. The proposed watermarking approach can hide visually recognizable patterns in images. The proposed approach is based on the discrete wavelet transform (DWT). This approach converts the color image from a RGB model to YCbCr representation, the Y channel is then decomposed into wavelet coefficients and the watermark is then hidden in the luminance component Y, because of the human eye is less sensitive to luminance in YCbCr space than other color channels in RGB space. Then the watermark is encrypted by Arnold transform algorithm. Afterwards, we embed watermark in the approximation coefficients of DWT of the host image by modifying least significant bit.

*A. Watermark Embedding Method*

The block diagram of the watermarking embedding is shown in Figure 2.The algorithm for embedding watermark in LL3 coefficients of the host image Y channel is described as follows:

Step 1: Convert RGB channels of a host image W into YCbCr channels

Step 2: Decompose the Y channel into a three-level wavelet pyramid structure with ten DWT sub bands. The coarsest sub band LL3 is taken as the target sub band for embedding watermarks.

Step 3: Save the signs of selection coefficients in a matrix sign.

Step 4: Quantize absolute values of selection coefficients.

Step 5: Scramble watermark W1 by Arnold transform algorithm for key times.

Step 6: Embedding watermark W1. For robustness, imperceptibility, and security, the watermark W is embedded in least significant bit that has smallest quantization errors.

Step 7: Effect matrix sign into the embedded coefficients.

Step 8: Reconvert YCbCr channels of a changed host image into RGB channels.

Step 9: A watermarked image W' is then generated by inverse DWT with all changed and unchanged DWT coefficients.

Step10: Save the key times of Arnold transformation, indexes of changed selection coefficients, and index of the embedded sub band as the authenticated key.

*B. Watermark Extraction Method*

Figure3 shows the watermark extraction method. The embedded watermark can be extracted using the stored authenticated key after wavelet decomposition of the watermarked image. The extracting process is described as follows:

Step 1: The RGB channels of the watermarked image are converted into YCbCr channels.
Step 2: Decompose the Y channel into ten DWT sub-bands.
Step 3: Re-fetch the stored authenticated key.
Step 4: Quantize absolute values of LL3 sub-band.

Step 5: Extract least significant bit of re-fetched key.
Step 6: re-scrambling watermark by inverse Arnold transform for key times.

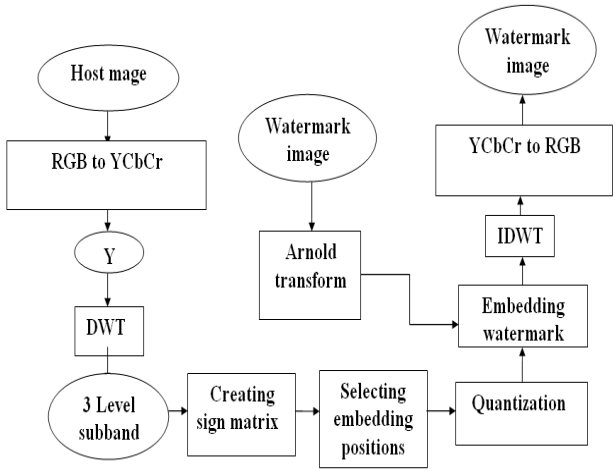

Fig. 2 Embedding procedure

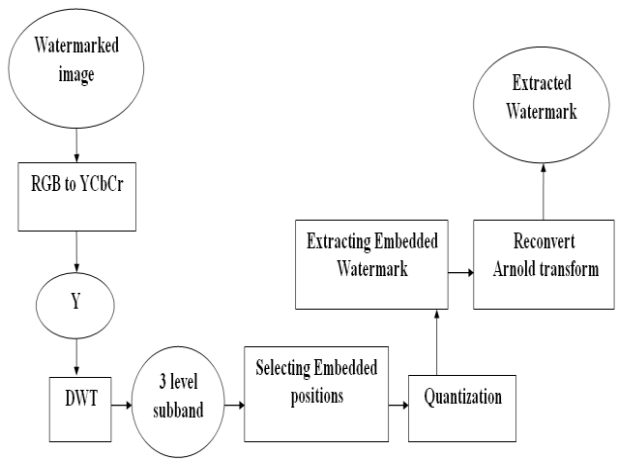

Fig.3 Extraction procedure

## VI. EXPERIMENTS AND RESULTS

We choose Three 256×256 famous images: Baboon, F16 and Arm, shown in Figure 4(a-c) were taken as the host images to embed a 30×30 binary watermark image, shown in Figure 4(d). For the entire test results in this paper, MATLAB R2007a software is used. Also for computing the wavelet transforms, 9-7 biorthogonal spline (Bspline) wavelet filter are used. Cause of use of B-spline function wavelet is that, B-spline functions, do not have compact support, but are orthogonal and have better smoothness properties than other wavelet functions [10].
In order to evaluate the quality of image, we use parameter peak value signal-to-noise ratio (PSNR) [11].

$$PSNR = 10\ \log_{10} \frac{255^2}{MSE}\ (dB) \qquad (4)$$

Where mean-square error (MSE) is defined as:

$$MSE = \frac{1}{mn} \sum_{i=1}^{m}\sum_{j=1}^{n} \left(h_{i,j} - h'_{i,j}\right)^2 \qquad (5)$$

Where $\{h_{i,j}\}$ and $\{h'_{i,j}\}$ are the gray levels of pixels in the host and watermarked images, respectively. The larger PSNR is the better the image quality. In general, a watermarked image is acceptable by human perception if its PSNR is greater than 30dBs. In other words, the correlation is used for evaluating the robustness of watermarking technique and the PSNR is used for evaluating the transparency of watermarking technique [12].
In order to evaluate the robustness of watermarking algorithm, the comparability between original watermark w and detected watermark w* is calculated with the formula hereinafter [5]:

$$NC = \frac{\sum_i \sum_j w_{ij} \times w_{ij}*}{\sum_i \sum_j w(i,j)^2} \qquad (6)$$

We also measure the similarity of original host image and watermarked images by the standard correlation coefficient[5]:

$$Correlation = \frac{\sum (x-x')(y-y')}{\sqrt{(x-x')^2}\sqrt{(y-x')^2}} \qquad (7)$$

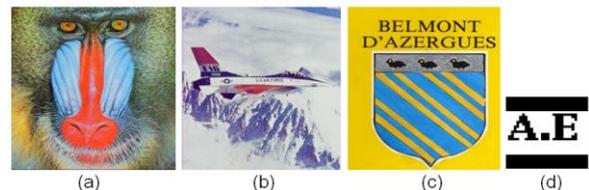

Fig.4 the host images for watermarking.
(a-c) Baboon, F16, and Arm. (d) Watermark image

### A. Imperceptibility Experiment

The obtained results of imperceptibility experiment show that the proposed algorithm satisfies watermark imperceptibility as well and improves results in [5,9]. The PSNR values of the watermarked images produced by the proposed approach are all greater than 84.5968 dB, NCs between original watermark images and extracted watermark images are all equal 1, and correlations between host images and watermarked images are all greater than 0.999.
Table 1 shows the imperceptibility results of proposed scheme.

### B. Robustness of watermark

In this part we measured the robustness of watermark against wavelet compression and cropping attacks.

*1) Robustness to wavelet Compression Attacks:* In this part we tested the watermark robustness against wavelet compression. We compressed the watermarked images with different thresholds with using wavelet compression. After that we extracted the watermark. We observed in all of these examinations, all PSNRs with threshold of less than or equal 7 were more than 84dB, and NCs were more than 0.80. The details of results are shown in Table 2. The obtained results of robustness experiment show that the proposed algorithm satisfies watermark robustness as well and improves results in [5,9].

2) *Robustness to Cropping Attacks:* In this part of our examination we crop three different parts of the watermarked images and then, we extract the watermark images from the cropped images. We can see all in these results the NCs are still more than 0.87. In table 3 are shown the extracted watermarks and the cropped parts of watermarked images.

TABLE I

THE IMPERCEPTIBILITY RESULTS OF PROPOSED SCHEME

| Image | PSNR(dB) | Extracted Watermark | Corr | NC | Error Bits% |
|---|---|---|---|---|---|
| F16 | 89.5101 | 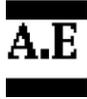 | 0.9997 | 1.00 | 0 |
| Arm | 88.0237 | 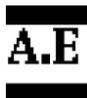 | 0.9997 | 1.00 | 0 |
| Baboon | 84.5968 | 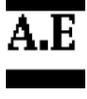 | 0.9996 | 1.00 | 0 |

TABLE IIIII

OBTAINED RESULTS OF WAVELET COMPRESSION ATTACKS WITH DIFFERENT COMPRESSION THRESHOLDS

| Threshold = 3.0 | | | | |
|---|---|---|---|---|
| Image | PSNR (dB) | Extracted Watermark | NC | Error Bits(%) |
| F16 | 89.5098 | 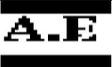 | 0.9933 | 0.3333 |
| Arm | 88.0234 | 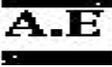 | 0.9933 | 0.3333 |
| Baboon | 84.5963 | 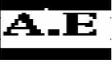 | 0.9955 | 0.2221 |
| Threshold = 5.0 | | | | |
| Image | PSNR (dB) | Extracted Watermark | NC | Error Bits(%) |
| F16 | 89.5092 | 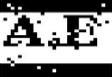 | 0.9287 | 3.5555 |
| Arm | 88.0229 | 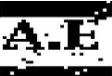 | 0.9198 | 4 |
| Baboon | 84.5951 | 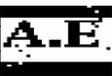 | 0.9732 | 1.333 |
| Threshold = 7.0 | | | | |
| Image | PSNR (dB) | Extracted Watermark | NC | Error Bits(%) |
| F16 | 89.5084 | 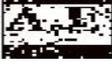 | 0.8060 | 9.6667 |
| Arm | 88.0221 | 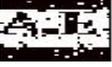 | 0.8417 | 7.8889 |
| Baboon | 84.5931 | 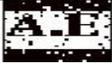 | 0.8863 | 5.6667 |

TABLE III

OBTAINED RESULTS OF CROPPING ATTACKS

| Robustne | Cropped Area 1 | | | |
|---|---|---|---|---|
| Image | Cropped image | Extracted Watermark | NC | Error Bits(%) |
| F16 | 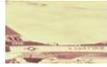 | 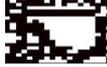 | 0.8769 | 44 |
| Arm | 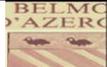 | 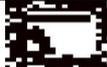 | 0.9250 | 44.33 |
| Baboon | 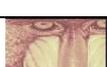 | 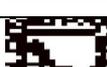 | 0.9112 | 43.77 |
| Cropped Area 2 | | | | |
| Image | Cropped image | Extracted Watermark | NC | Error Bits(%) |
| F16 | 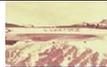 | 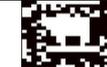 | 0.9088 | 42.22 |
| Arm | 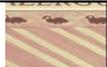 | 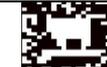 | 0.9014 | 38.66 |
| Baboon | 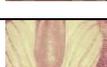 | 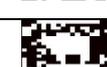 | 0.9157 | 40.55 |
| Cropped Area 3 | | | | |
| Image | Cropped image | Extracted Watermark | NC | Error Bits(%) |
| F16 | 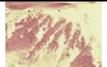 | 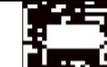 | 0.8782 | 45.22 |
| Arm | 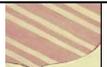 | 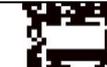 | 0.8707 | 44.33 |
| Baboon | 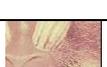 | 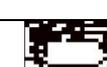 | 0.8704 | 44.77 |

VII. CONCLUSIONS

In this paper we propose an efficient watermarking algorithm for digital image. The proposed algorithm has good security and cannot be easily attacked. We embedded a binary image watermark into color images and we found that after image-processing attacks especially the wavelet compression attacks, the results are more improved than [5,9].

In this study, we converted a host image into YCbCr color space, and then, the Y channel was decomposed into wavelet coefficients. Then, by using LSB insertion technique, the watermark was encrypted with Arnold transform and embedded into the least significant bit of host coefficients in the approximation coefficients subband. Extra imperceptibility and robustness of watermarking against wavelet compression attacks compared with [5,9] were observed in our study. Moreover, the proposed approach had no need of the original host image to extract watermarks.